\documentclass[galaxies,article,accept,moreauthors,pdftex,10pt,a4paper]{mdpi}

\usepackage{epstopdf}

\firstpage{1}
\pubvolume{xx}
\issuenum{1}
\articlenumber{1}
\pubyear{2018}
\copyrightyear{2018}
\history{Received: date; Accepted: date; Published: date}

\Title{A catalog of photometric redshift and the distribution of broad galaxy morphologies}

\Author{Nicholas Paul$^1$, Nicholas Virag$^1$ and Lior Shamir$^1$*}

\AuthorNames{Nicholas Paul, Nicholas Virag and Lior Shamir}

\address{%
$^{1}$ \quad Department of Math and Computer Science, Lawrence Technological University, Southfield, MI 48075 
}

\corres{Correspondence: lshamir@mtu.edu; Tel.: +1-248-204-3512}

\abstract{We created a catalog of photometric redshift of $\sim$3,000,000 SDSS galaxies annotated by their broad morphology. The photometric redshift was optimized by testing and comparing several pattern recognition algorithms and variable selection strategies, trained and tested on a subset of the galaxies in the catalog that had spectra. The galaxies in the catalog have i magnitude brighter than 18 and Petrosian radius greater than 5.5''. The majority of these objects are not included in previous SDSS photometric redshift catalogs such as the {\it photoz} table of SDSS DR12. Analysis of the catalog shows that the number of galaxies in the catalog that are visually spiral increases until redshift of $\sim0.085$, where it peaks and starts to decrease. It also shows that the number of spiral galaxies compared to elliptical galaxies drops as the redshift increases. The catalog is publicly available at \url{https://figshare.com/articles/Morphology_and_photometric_redshift_catalog/4833593}.}


\begin{document}

\section{Introduction}
\label{introduction}

Most galaxies can be broadly separated into two morphological types -- spiral and elliptical \citep{hubble1926extragalactic}. Galaxy Zoo \citep{lintott2011galaxy} was the first attempt to analyze the distribution of a large number of spiral and elliptical galaxies in the local universe. Using the power of crowdsourcing, it provided morphological classifications of nearly 900,000 galaxies. Subsets of ``clean'' and ``superclean'' datasets were used to deduce the distribution of elliptical and spiral galaxies in the local universe \citep{bamford2009galaxy,schawinski2009galaxy,skibba2009galaxy}. 

A more recent catalog of galaxy morphology is the catalog of $\sim$3,000,000 Sloan Digital Sky Survey (SDSS) galaxies \citep{kuminski2016computer} classified automatically using machine learning \citep{shamir2009automatic,kuminski2014combining}. While the catalog is large, it is limited in the sense that the vast majority of the galaxies in that catalog do not have spectroscopic data.

Photometric redshift (photo-z) plays a vital role in the study of astronomy and cosmology. Spectroscopic measurements of millions of celestial objects is technically daunting and expensive compared to photometric measurements. The redshift can be estimated from the photometric measurements, and that estimation is often sufficient for many applications involving statistical analysis of a large population of astronomical objects \citep{oyaizu2008galaxy}. Clearly, photometric measurements can provide more redshift estimates per unit telescope time compared to spectroscopic measurements \citep{hildebrandt2010phat}. Therefore, during the past decade significant efforts have been aimed toward developing photometric redshift estimation methods, most of them can be classified into two types: template-fitting and empirical methods \citep{zheng2012review}.  

Empirical methods use celestial objects with known spectroscopic redshift as training data to estimate the redshift based on patterns in the photometric measurements. The performance of empirical methods is limited to the range of the spectroscopic redshift of the samples in the training set. Template-based methods estimate the photometric redshift by using spectral templates. The estimation is accomplished by selecting the SED from a library of templates such that the SED best reproduces the observed fluxes in the broadband filters \citep{oyaizu2008galaxy}. These methods are preferred when exploring new regimes, while empirical methods are preferred when large training sets with spectroscopic redshift are available \citep{zheng2012review}. In general, empirical methods provide better accuracy \citep{zheng2012review}. For detailed discussion of photometric redshift techniques see reviews \citep{zheng2012review,ball2010data}.

In some cases hybrid methods that combine empirical and template approaches can improve the accuracy of the photometric redshift reconstruction. Examples of empirical methods include predication trees and random forests \citep{kind2013tpz,carliles2010random}, Polynomial Fitting \citep{connolly1995slicing}, the Nearest Neighbour Polynomial (NNP) technique \citep{cunha2009estimating}, Decision Trees \citep{gerdes2010arborz}, Artificial Neural Networks \citep{collister2007megaz,vanzella2004photometric}, and Support Vector Machines \citep{wadadekar2005estimating}. Gaussian Process Regression (GPR) can provide competitive results compared to ANN and least-square fitting methods \citep{way2009new}. Some studies showed improved prediction accuracy using the combination of templates and magnitude priors \citep{schmidt2013improved}. The inclusion of near IR magnitude and angular size also showed significant contribution to the accuracy of the photometric redshift prediction \citep{gomes2017improving}. The combination of morphological and photometric variables have shown to increase the accuracy, especially in cases where fewer bands are available \citep{soo2017morpho}. 

Examples of photometric redshift catalogs include the catalog of $\sim10^6$ SDSS DR4 objects with redshift values in the range of $0.4 < z < 0.7$ \citep{collister2007megaz}, and the catalog of SDSS DR9 galaxies, in which an artificial neural network was used \citep{brescia2014catalogue}. The ANNz2 artificial neural network \citep{sadeh2016annz2} was used to create a photometric redshift catalog of $\sim3.9\cdot10^8$ for the Kilo-Degree Survey Data Release 3 \citep{bilicki2017photometric}. Another large catalog is contains the photometric redshift catalog of about $\sim2\cdot10^8$ galaxies from SDSS DR12, with redshift range of $0<z<$0.8  \citep{beck2016photometric}. 

Here we test several machine learning algorithms and sets of photometric variables for the purpose of photometric redshift estimation, and apply the method to create a catalog of $\sim$3,000,000 galaxies that have information about their broad morphology and their photometric redshift. That information can be used to profile the distribution of broad morphology of galaxies in different redshift ranges.


\section{Data}
\label{data}

The initial data were taken from the catalog of $\sim3\cdot10^6$ SDSS galaxies separated by their broad morphology into elliptical and spiral galaxies \citep{kuminski2016computer}, all taken from SDSS DR8 \citep{rykoff2014redmapper}. The vast majority of these DR8 galaxies do not have photometric redshift computed through previous catalogs. For instance, a joint query with the redshift catalog of $\sim2\cdot10^8$ SDSS DR12 galaxies \citep{beck2016photometric} only includes 827,591 of the galaxies in the catalog, which is merely about 27.6\%.

\begin{figure}
  \centering
  \includegraphics[width=1.0\textwidth]{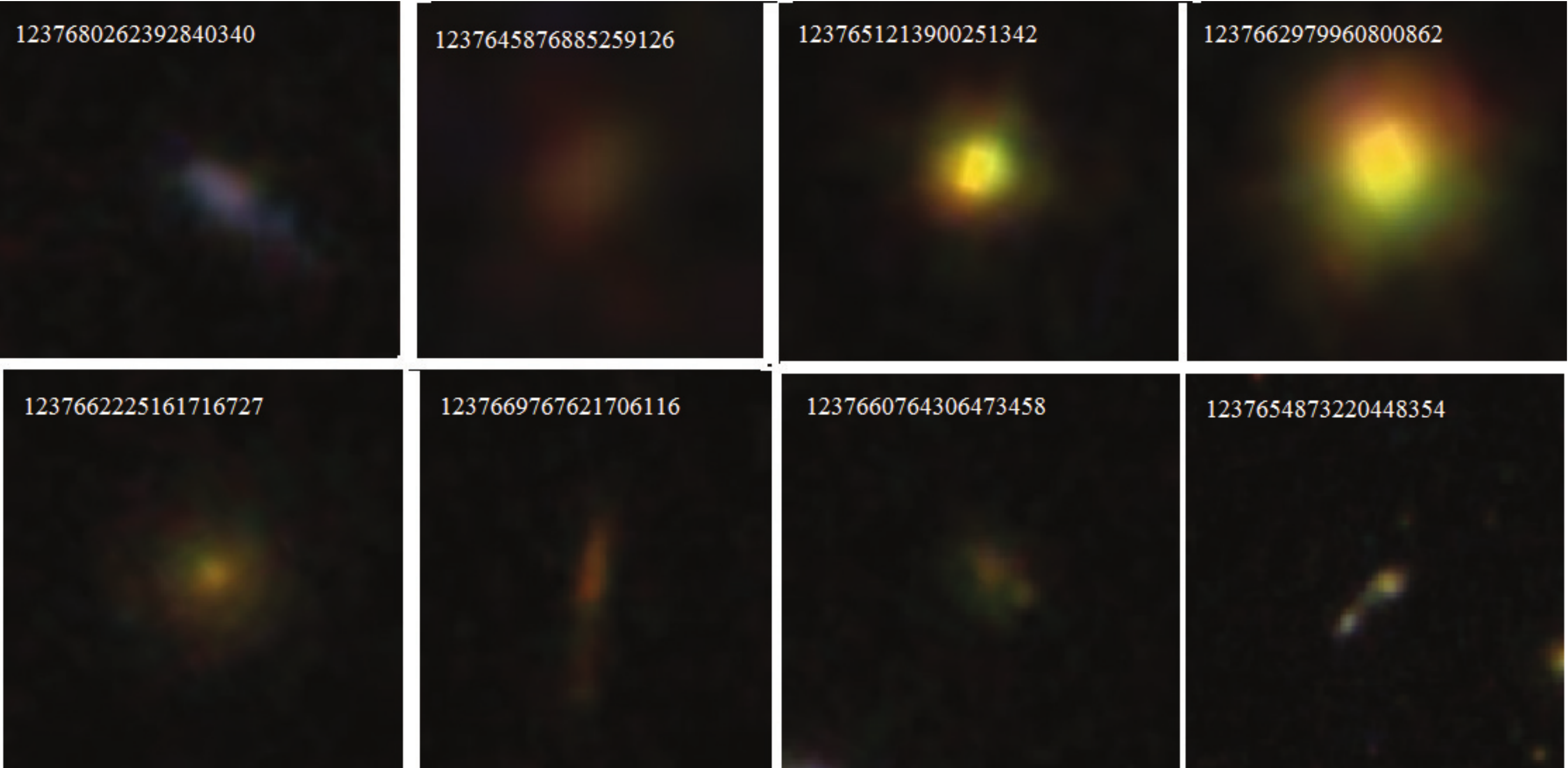}
  \caption{Examples of objects identified as galaxies by DR12 PhotoObjAll table, but are not included in the photoz table of DR12.}
  \label{dr12}
\end{figure}

The reason for the exclusion of these galaxies from the {\it photoz} table \citep{beck2016photometric} of SDSS DR12 could be that the galaxies in the photoz table are objects identified as galaxies by all primary photometric measurements included in the GalaxyTag view \citep{beck2016photometric}, while the objects in the catalog of broad morphology were selected by using the ``type'' field of the PhotoObjAll table, and then filtered by applying further analysis of the image based on the morphology of the object. Therefore, many objects that have an object type ``galaxy'' (type=3) in the PhotoObjAll table and are included in the catalog of broad morphology might not have been included in the set objects included in the photoz table. For instance, DR12 has 48,528,684 objects with model i magnitude smaller than 19 and identified to have the type ``galaxy'', while just 11,761,054 of these objects are included in the photoz table. Figure~\ref{dr12} shows examples of objects identified as galaxies in SDSS DR12 PhotoObjAll, but are not included in the photoz table of DR12. 

The catalog also has the certainty of each galaxy to belong in each of the broad morphological classes, and the threshold can be used to control the consistency of the subset \citep{kuminski2016computer}. The certainty values are used in a similar fashion to the way the degree of agreement between human annotators is used in Galaxy Zoo \citep{lintott2011galaxy}. By using certainty threshold of 0.54 for the spiral galaxies and 0.8 for the elliptical galaxies, the catalog contains $\sim9\cdot10^5$ spiral galaxies and $\sim6\cdot10^5$ elliptical galaxies with consistency of $\sim$98\% with the Galaxy Zoo debiased ``superclean" dataset \citep{lintott2011galaxy}, as thoroughly described in \citep{kuminski2016computer}. All galaxies are bright (i magnitude brighter than 18) and large (Petrosian radius measured in the r band larger than 5.5''), and therefore allow the identification of the morphologies of the galaxies in the catalog while excluding small and faint objects that their morphology cannot be identified.

\begin{figure}
  \centering
  \includegraphics[width=1.0\textwidth]{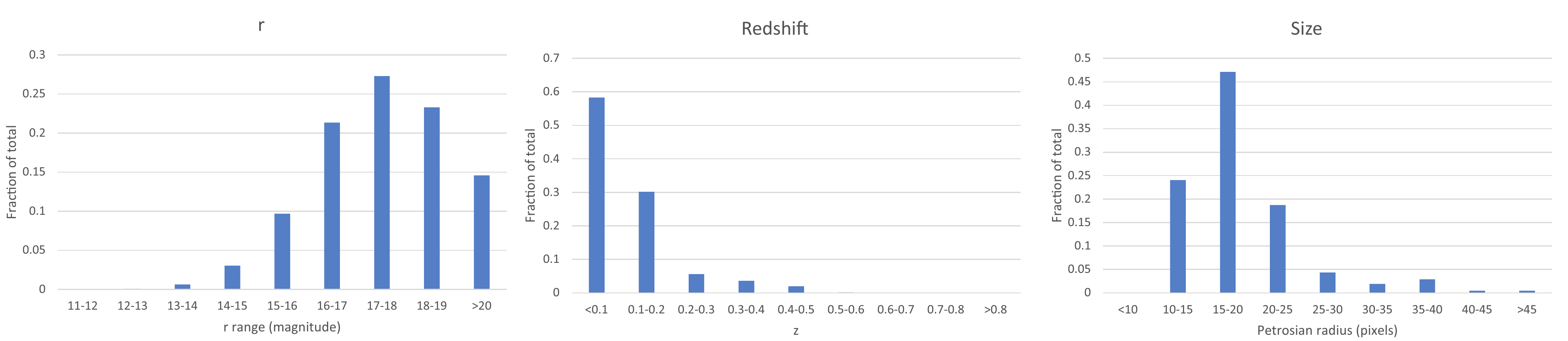}
  \caption{Distribution of the redshift, r magnitude, and Petrosian radius measured on the r band of the 115,359 galaxies that have spectra. These galaxies were used for training and testing the photometric redshift algorithms.}
  \label{catalog_spec_distribution}
\end{figure}

The source code used to create the catalog is also publicly available \citep{shamir2013wnd}. Figure~\ref{catalog_spec_distribution} shows the distribution of the r model magnitude, the Petrosian radius measured in the r band, and the distribution of the redshift among 115,359 galaxies included in the catalog that also had spectra.

A subset of 20,000 galaxies that have spectra was used for training and testing the algorithm. Naturally, these galaxies need to have spectra so that the predicted photometric redshift can be compared to their spectroscopic redshift to deduce the efficacy of the algorithm. The photometric information was taken from the PhotoObjAll table of SDSS DR8. 

\begin{figure}
  \centering
  \includegraphics[width=1.0\textwidth]{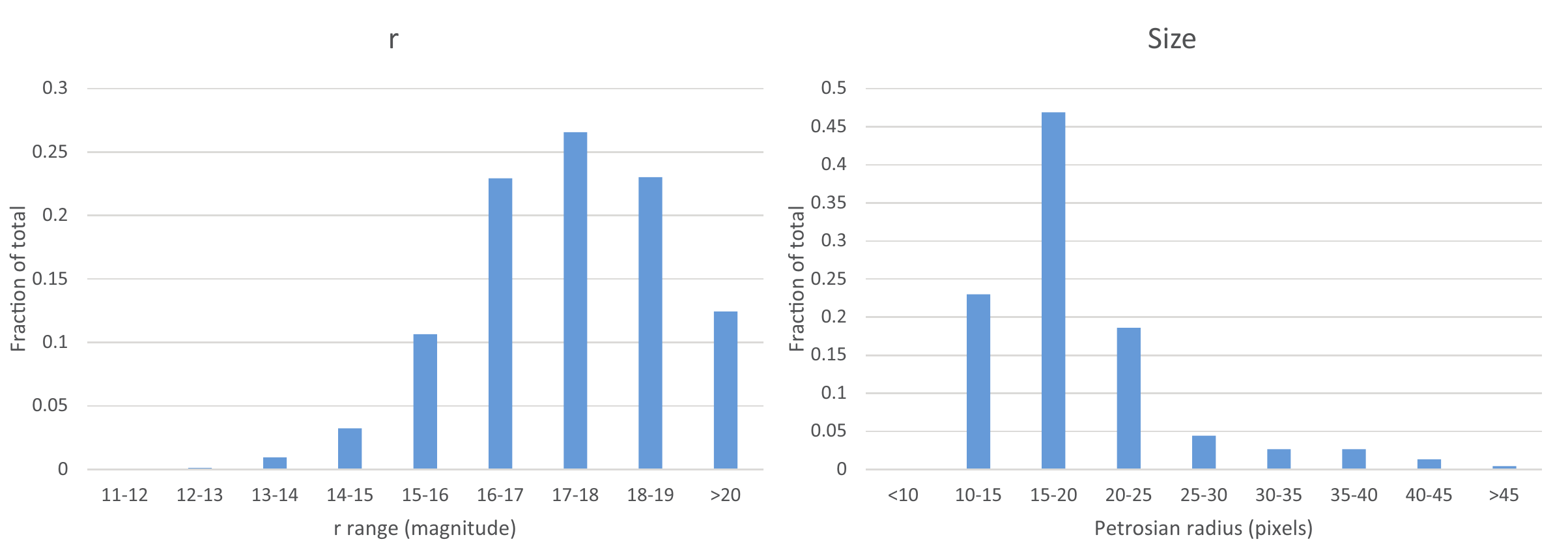}
  \caption{Spectoroscopic redshift distribution of the galaxies in the catalog of broad galaxy morphology.}
  \label{histogram_catalog}
\end{figure}

The purpose of the set of galaxies with spectra is to train and test a model to estimate the redshift of the galaxies in the catalog of SDSS galaxies with broad morphology classification \citep{kuminski2016computer}. For that purpose, the set of galaxies with spectra that are used for training and testing needs to be as similar as possible to the entire population of galaxies in the catalog it aims at analyzing. Figure~\ref{histogram_catalog} shows the distribution of the magnitude and size of the galaxies in the catalog of galaxies with broad morphological classification. As the figure shows, the brightness and size of the galaxies in the catalog is very similar to the brightness and size of the galaxies in the training set.

Selecting training samples from the same set of galaxies that need to be classified might lead to performance evaluation that reflects the sample from which the galaxies were taken, and not necessarily the entire set of SDSS galaxies, which also includes small and faint galaxies. Also, the training set contains galaxies that were selected as spectroscopic targets, and are therefore not necessarily a random representation of SDSS general galaxy population. However, since the purpose of the algorithm is to estimate the redshift only for galaxies within that sample, higher accuracy can be achieved if the population of the samples in the training set is similar to the population of the samples that will be classified with the machine learning system. The galaxies in both the catalog and the training set are galaxies that were selected using the same criteria, and Figures~\ref{catalog_spec_distribution} and~\ref{histogram_catalog} show that the population of galaxies in the catalog is similar to the population of galaxies included in the training set. The training set does not include random galaxies from SDSS spectroscopic sample, but just galaxies that are part of the catalog, and their population is similar to the galaxy population in the catalog. While the solution is expected to achieve poor performance for galaxies outside of that sample, it is designed specifically for a certain catalog.

\section{Methods}
\label{methods}

\subsection{Pattern recognition algorithms}

Several supervised machine learning algorithms were tested, and the performance was evaluated to identify the algorithms that demonstrated the highest efficacy. These algorithms included Simple Linear Regression, MultiLayer Perceptron \citep{gardner1998artificial}, M5P \citep{quinlan1992learning}, ZeroR, Decision Table \citep{witten2005data}, and Random Forest \citep{Breiman2001}. Because the photometric redshift is a continuous value and not a crisp class, suitable algorithms need to be able to perform a regression and compute a continuous value as their output.

The \textit{Logistic Regression} algorithm predicts a multi-dimensional point by minimizing its squared error. \textit{MultiLayer Perceptron} builds a multi-layer neural network of weighted perceptron nodes. Each node receives several input values, and ``fires'' a value to the next layer if the results of the function (called ``activation function'') using these input values as parameters reaches a certain threshold weight \citep{belue1995determining}. The weights in the nodes are optimized by running the training samples through the network iteratively, and adjusting the weights based on the results of the training samples. Because the output layer contains multiple perceptrons, their values can be interpolated to provide a continuous value.

The \textit{M5P} algorithm implements M5 model trees and rules \citep{quinlan1992learning}. Because in the M5 model each leaf is a linear regression function, it is suitable for predicting continuous values rather than a crisp class. \textit{ZeroR} is a simple classifier that makes a prediction based on the frequency of the output variable in the training set. A \textit{Decision Table} \citep{witten2005data} is a rule-based method that uses a frequency table to make a prediction. The frequency table is built based on the frequency of the features in the training samples, and their distribution in different ranges \citep{witten2005data}. The tree-based \textit{Random Forest} algorithm \citep{Breiman2001} builds a classifier using a large number of random decision tree classifiers. Each decision tree is created randomly such that each node is a different feature, and the decision is made based on the value of that feature in a specific given test sample, until reaching the leaf that is assigned with an output value. These trees are used to create an ensemble classifier such that each tree is a classifier, and the output is determined by an interpolation of the results of all decision trees. Because each decision tree provides an output, the high number of outputs being interpolated makes the method suitable also for the prediction of continuous values. The implementation of the algorithm was taken from the Weka open source machine learning toolbox \citep{witten1999weka}.

\subsection{Variable selection}
\label{variable_selection}

Feature selection in a multidimensional environment is a complex task that often requires heuristics or assumptions, which can then be tested empirically. Several different methods were used for variable selection, including hand-crafted variable selection and automatic statistical selection of the variables. The hand-crafted set of variables is the variables used in \citep{brescia2014catalogue} to compute the photometric redshift of SDSS DR9 objects. These variables are listed in Table~\ref{Variable_sets}, and a short description of each variable can be found in Table~\ref{variable_description}.

\begin{table}
\caption{Variable sets}
\label{Variable_sets}
\begin{tabular}{|p{1.7cm}|p{6cm}|p{0cm}|}
\hline
Name      & Variables \\                                                                                                                                                                                                                                                                                                    \hline
\citep{brescia2014catalogue}     & ra, dec, g, r, i, z, psfMagErr u, psfMagErr g, psfMagErr r, psfMagErr i, psfMagErr z, extinction u, extinction g, extinction r, extinction i, extinction z, u-g, u-r, u-i, u-z   \\ 
KBest13     & deVMag\_g, deVMag\_r, dered\_g, dered\_u, expMag\_g, expMag\_r, fiberMag\_u, g, petroMag\_g, photozcc2, photozd1, u                                                                                                                                                                                              \\ 
KBest21  &  deVMag\_g,deVMag\_r, deVMag\_u, dered\_g, dered\_r, dered\_u, expMag\_g, expMag\_r, fiberMag\_g, fiberMag\_u, g, petroMag\_g, petroMag\_r, photozcc2, photozd1, psfMag\_g, psfMag\_u, r, u  \\ 
KBest30  &  deVMag\_g, deVMag\_i, deVMag\_r, deVMag\_u, dered\_g, dered\_r, dered\_u, expMag\_g, expMag\_i, expMag\_r, expPhiErr z, fiberMag\_g, fiberMag\_u, g, isoA\_u, isoBGrad\_u, isoB\_u, isoCocl\_u, isoPhi\_u, isoRowcGrad\_u, petroMag\_g, petroMag\_r, petroMag\_u, photoscc2, photozd1, psfMag\_g, psfMag\_u, r, u \\ 
KBestMod &  deVMag\_g, deVMag\_r, deVMag\_u, dered\_g, dered\_r, dered\_u, expMag\_g, expMag\_r, fiberMag\_g, fiberMag\_u, g, petroMag\_g, petroMag\_r, psfMag\_g, psfMag\_u, ra, dec, i, r, u, u-g, g-r, r-i, i-z   \\ 
\hline
\end{tabular}
\end{table}



Another method of variable selection is based on computing the variable's analysis of variance (ANOVA) F-value, and then selecting the highest rated features \citep{py-ml2016raschka}. That was done by using the python library \textit{scikit-learn}\footnote{http://scikit-learn.org/stable/index.html}. The function \texttt{sklearn.feature\_sclection.SelectKBest} takes a dataset and a comparison function to choose the most informative features. For the comparison function, we used the ANOVA F-value computation function \texttt{sklearn.feature\_selection.f\_classif}. These variables are selected automatically, and therefore some of the variables might not necessarily have a straightforward physical explanation, but in the context of the database can provide useful patterns when used in combination with other variables. We chose the top $13$ (KBest13), $21$ (KBest21), and $31$ (KBest31) variables and ran the random forest algorithm on each subset of variables.

Table~\ref{tab:kbest-size} shows the mean absolute error when using the different feature sets. The mean absolute error is defined by $\frac{\Sigma_{i=1}^N |Z_{i,p}-Z_{i,s}|}{n}$ , where $Z_{i,p}$ is the photometric redshift of galaxy {\it i}, $Z_{i,s}$  is the spectroscopic redshift of galaxy {\it i}, and N is the total number of galaxies in the test samples.  That process is repeated 10 times using 10-fold cross-validation test strategy, meaning that the test is performed 10 times such that in each run a different set of 10\% of the samples are used for testing, and the remaining 90\% for training.

\begin{table}
\caption{Mean absolute error for three different feature sets when using the random forest algorithm and the 10-fold cross-validation test strategy.}
\label{tab:kbest-size}
\begin{tabular}{|l|l|}
\hline
Feature Set & Mean Abs Error \\ 
\hline
KBest13     & 0.00749        \\ 
KBest21     & 0.00741        \\ 
KBest30     & 0.00748        \\ 
\hline
\end{tabular}
\end{table}

As Table \ref{tab:kbest-size} shows, using $21$ variables provided the best performance, and the larger feature set did not lead to better accuracy, although the difference in performance when using different sets of variables is small. All experiments were performed with a standard 10-fold cross-validation testing strategy.

Finally, the set of KBest21 variables was enhanced with the four color variables, that are not included in SDSS and were therefore not analyzed by ANOVA, creating the KBestMod variable set. That variable set included the following variables: deVMag\_g, deVMag\_r, deVMag\_u, dered\_g, dered\_r, dered\_u, expMag\_g, expMag\_r, fiberMag\_g, fiberMag\_u, g, petroMag\_g, petroMag\_r, psfMag\_g, psfMag\_u, ra, dec, i, r, u, u-g, g-r, r-i, i-z.



\subsection{Performance evaluation}

In order to evaluate the performance we used three metrics of the performance: The mean absolute error, the root square error, and the normalized error. The root mean square error (RMSE) is defined by Equation~\ref{root_error}


\begin{equation}
RMSE = \sqrt{\frac{1}{N}  \Sigma_{i=1}^{N}(Z_{spec}-Z_{phot})^2 }.
\label{root_error}
\end{equation}

The normalized $Z$ error $\Delta Z_{norm}$ is defined by Equation~\ref{normalize}.

\begin{equation}
\Delta Z_{norm} = \frac{Z_{spec} - Z_{phot}}{1 + Z_{spec}}
\label{normalize}
\end{equation}

and the mean normalized error $\overline{\Delta Z_{norm}}$ is the mean $\Delta Z_{norm}$ of the test galaxies.


Table~\ref{tab:feature-sets} shows the performance for the different variable sets when using the random forest classifier. As the table shows, the \texttt{KBestMod} feature set performs better than the other feature sets for all three performance metrics.

\begin{table}
\caption{Performance of random forest on 20,000 galaxies with different feature sets}
\label{tab:feature-sets}
\begin{tabular}{|l|r|r|r|}
\hline
\multicolumn{1}{|c|}{\textbf{Feature set}} & \textbf{Mean Abs. Err.} & \textbf{Root Err.} & \textbf{$\overline{\Delta Z_{norm}}$} \\ 
\hline
Brescia et al. \cite{brescia2014catalogue}  & 0.00747 & 0.01492 & 0.00246 \\ 
KBest13 & 0.00749 & 0.01728 & 0.00305 \\ 
KBest21  & 0.00741 & 0.01561 & 0.00259 \\ 
KBestMod & 0.00617 & 0.01294 & 0.00222 \\ 
\hline
\end{tabular}
\end{table}

Additionally, the random forest algorithm performs better than other algorithms. Table~\ref{tab:algorithms} shows the performance of the Simple Linear Regression, MultiLayer Perceptron, M5P, ZeroR, Decision Table, and Random Forest machine learning algorithms when using the \texttt{KBestMod} feature set.

\begin{table}
\caption{The 10-fold cross-validation performance of the different machine learning methods on a set of 20,000 galaxies using the \texttt{KBestMod} feature set.}
\label{tab:algorithms}
\begin{tabular}{|l|r|r|r|r|}
\hline
\multicolumn{1}{|c|}{\textbf{Algorithm}} & \multicolumn{1}{c|}{\textbf{Mean Abs. Err.}} & \multicolumn{1}{c|}{\textbf{Root Err.}} & \multicolumn{1}{c|}{\textbf{$\overline{\Delta Z_{norm}}$}} & \multicolumn{1}{c|}{\textbf{$\sigma(\overline{\Delta Z_{norm}})$  }}  \\ 
\hline
Simple Linear Regression    & 0.00621 & 0.01321 & 0.01273 & 0.0347 \\ 
MultiLayer Perceptron & 0.00617 & 0.01313 & 0.01041 & 0.0159 \\ 
M5P                   & 0.00611 & 0.01288 & 0.00562 & 0.019 \\ 
ZeroR                 & 0.00616 & 0.01304 & 0.06581 & 0.0719 \\ 
Decision Table    & 0.00616 & 0.01301 & 0.01586 & 0.0231 \\ 
Random Forest  & 0.00617 & 0.01294 & 0.00222 & 0.0107 \\ 
\hline
\end{tabular}
\end{table}

As the table shows, the random forest algorithm provided the best performance, with $\overline{\Delta Z_{norm}}$ of $\sim0.0022$, lower than any of the other algorithms. The standard deviation of the normalized error is also the lowest when using the random forest classifier (0.0107). The median absolute deviation of the random forest classifier is $\sim$0.056. M5P produced a marginally better mean absolute error and root square error of $0.0061$ and $0.0128$, respectively. Simple Linear Regression provided the worst performance, with a mean absolute error of $0.0062$ and root absolute error of $0.0132$.    


It should be mentioned that the performance figures provided in Table~\ref{tab:algorithms} reflect the performance on the galaxy sample of the catalog described in Section~\ref{data}, but not the entire galaxy population in SDSS.


The performance of the algorithm was also compared to the performance of the redshift estimation using Multilayer Perceptron, as well as the photometric redshift methods based on the Nearest Neighbour Polynomial (NNP), and the neural network algorithms CC2 and D1 \citep{oyaizu2008galaxy}. Table~\ref{comparison} shows that the photometric redshift estimation based on the random forest is favorably comparable to other methods, and therefore can provide a solution to the estimation of the redshift of the galaxies in the catalog.

\begin{table}
\caption{Error of the different photometric redshift algorithms in different redshift ranges.}
\label{comparison}
\begin{tabular}{|l|l|l|l|}
\hline
Measurement      & z $<$ 0.115   & 0.115 $ \leq z  $       & 0.117 $ \leq z$ \\ 
                           &                         & $ z < $ 0.177   & $ z < $ 0.345 \\ 
\hline 
$\overline{\Delta z_{norm}}$ Random Forest             &  0.002       & 0.002    & 0.003 \\ 
Mean Abs. Err. Random Forest         &  0.006       & 0.006    & 0.009 \\ 
Root Err. Random Forest         &  0.008       & 0.009    & 0.014 \\ 
$\overline{\Delta z_{norm}}$ Multilayer Perceptron     &   0.02       &  0.01    & 0.08 \\ 
Mean Abs. Err. Multilayer Perceptron &   0.03       &  0.02    & 0.36 \\ 
Root Abs. Err. Multilayer Perceptron &   0.044    &  0.04   & 0.41 \\ \hline
$\overline{\Delta z_{norm}}$ Err. NNP   &   0.062      &  0.108   & 0.16 \\ 
Mean Abs. Err. NNP                   &   0.067      &  0.118   & 0.19 \\ 
Root Err. NNP                   &   0.082      &  0.128   & 0.21 \\ 
$\overline{\Delta z_{norm}}$ Err. CC2    &   0.014      &  0.015   & 0.04 \\ 
Mean Abs. Err. CC2                   &   0.015      &  0.017   & 0.05 \\ 
Root Err. CC2                   &   0.058      &  0.058   & 0.13 \\ 
$\overline{\Delta z_{norm}}$ Err. D1                   &   0.024      &  0.018   & 0.06 \\ 
Mean Abs. Err. D1                    &   0.026      &  0.021   & 0.11 \\ 
Root Err. D1                    &   0.148      &  0.058   & 0.27 \\

\hline
\end{tabular}
\end{table}

\begin{figure}
  \centering
  \includegraphics[scale=0.45]{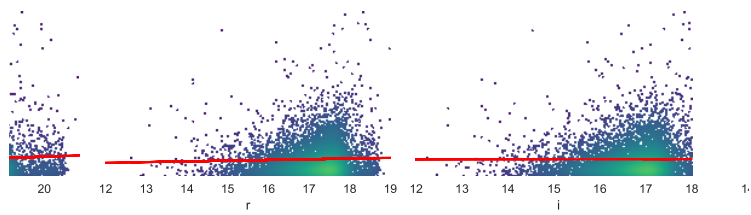}
  \caption{Absolute error when using the random forest classifier and different magnitude ranges.  The red lines show the least square error linear regression.}
  \label{mag_breakdown}
\end{figure}

\begin{figure}
  \centering
  \includegraphics[scale=0.45]{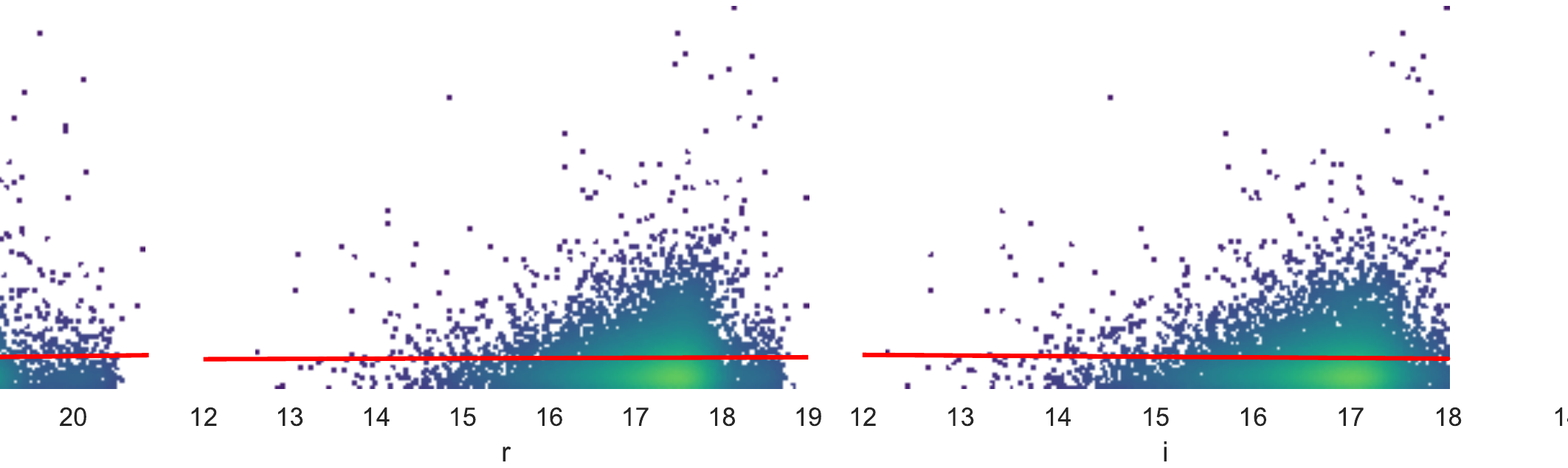}
  \caption{Normalized error when using the Random Forest classifier and different magnitude ranges.}
  \label{mag_breakdown_normalized}
\end{figure}

Photometric redshift accuracy can change in different color ranges. Figures~\ref{mag_breakdown} and~\ref{mag_breakdown_normalized} show the change in the absolute and normalized error, respectively, at different magnitudes in the different bands. The red lines show the least square error linear regression.

\begin{figure}
  \centering
  \includegraphics[scale=0.45]{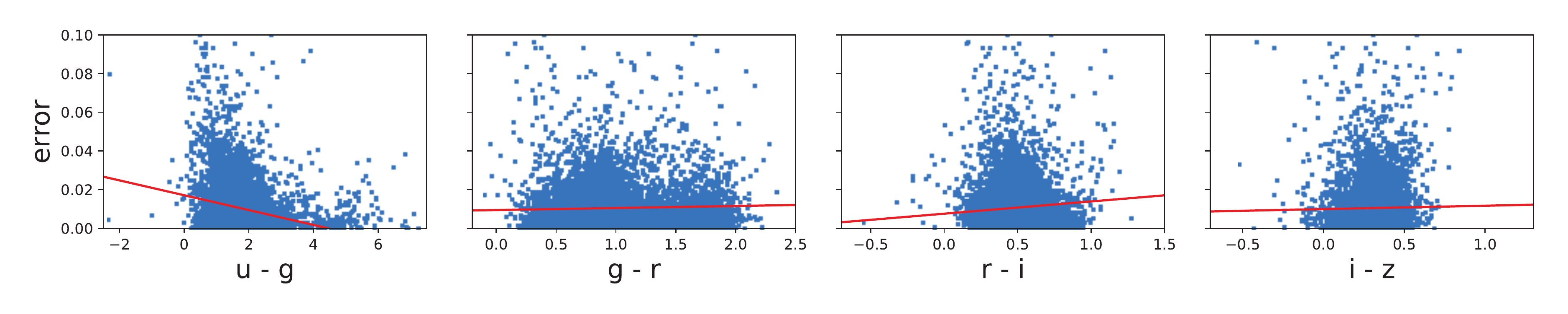}
  \caption{Absolute error when using the Random Forest classifier and different color ranges.}
  \label{color_error}
\end{figure}

\begin{figure}
  \centering
  \includegraphics[scale=0.45]{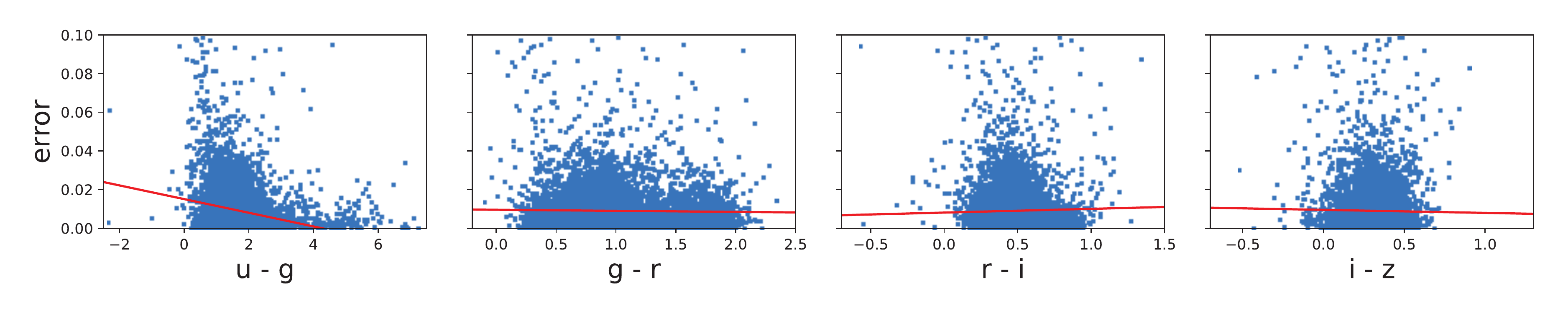}
  \caption{Normalized error when using the Random Forest classifier and different color ranges.}
  \label{color_error_norm}
\end{figure}

Figures~\ref{color_error} and~\ref{color_error_norm} show the absolute and normalized error, respectively, as a function of the color.

The figures show that the error increases when the objects get dimmer. The correlation between the error and the magnitude can be expected as the dimmer galaxies also tend to have higher redshift, and the error expected to increase as the redshift gets higher. The threshold of 0.54 was selected based on previous experiments by comparing it to the Galaxy Zoo ``superclean'' samples \citep{kuminski2016computer}. Any threshold higher than 0.54 does not increase the consistency of the dataset substantially. 

As Figures~\ref{color_error} and~\ref{color_error_norm} show, the error also tends to decrease slightly when the galaxies are bluer. That can be explained by the observation that spiral galaxies in the catalog tend to have lower redshift. Since spiral galaxies also tend to be bluer than elliptical galaxies, bluer galaxies in the catalog have lower redshift, and therefore their estimated photometric redshifts have a lower error. 

\begin{figure}
  \centering
  \includegraphics[scale=1.0]{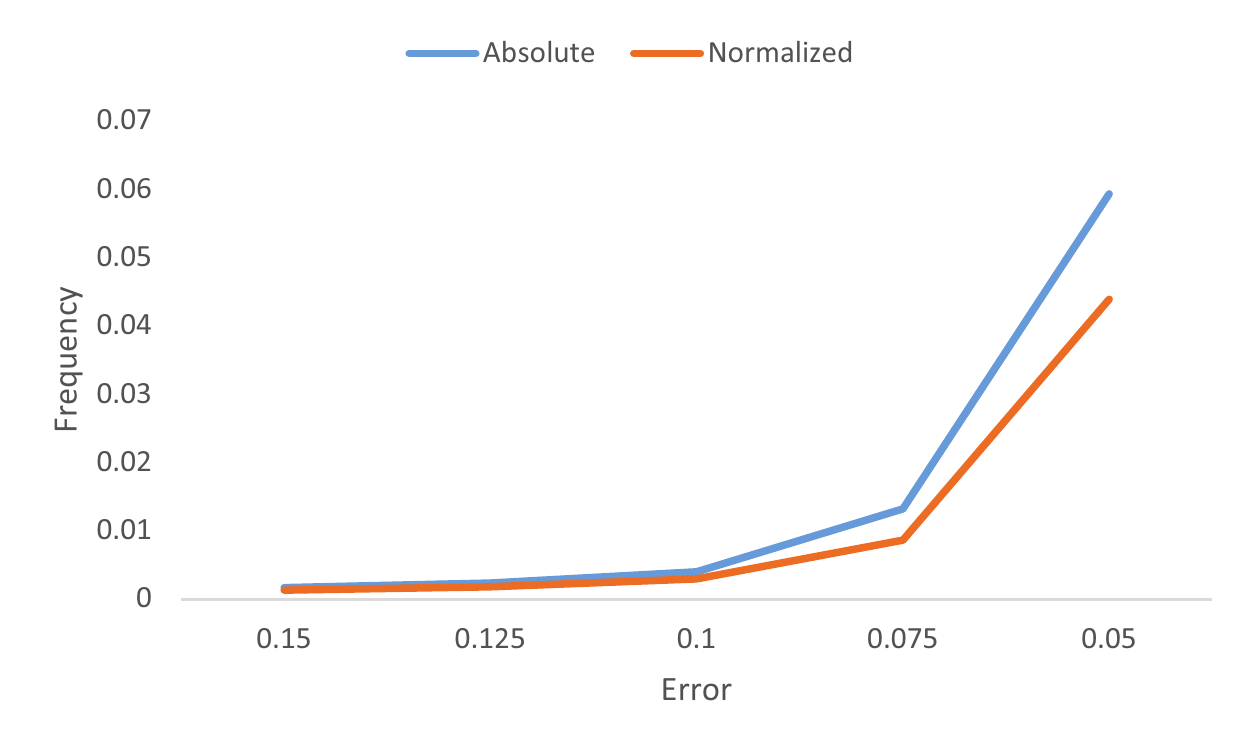}
  \caption{Frequency of galaxies as a function of the normalized and absolute error.}
  \label{outliers}
\end{figure}

Figure~\ref{outliers} shows the frequency of galaxies as a function of the normalized and absolute error. As the figure shows, catastrophic outliers with error of more than 0.15 are very rare, and are less than 0.2\% of the cases.   

\subsection{Dependence on the size of the training set}

The performance of machine learning algorithms is heavily dependent on the size of the dataset on which they are trained, and larger training sets normally lead to improved performance of the algorithm. However, the accuracy does not grow with the size of the training set in a linear fashion, and at a certain point it is expected that increasing the size of the training set has a negligible contribution to the performance of the machine learning algorithm \citep{christopher2006pattern}.

To determine the effect of the size of the training set on the performance, the random forest algorithm and the \texttt{KBestMod} feature set were used with several training set sizes ranging from $1,000$ to $20,000$ galaxies. The results of these experiments are shown in Table \ref{tab:training-size}. As the table shows, the accuracy of the algorithm improves as the size of the dataset gets larger, but the improvement in the $\overline{\Delta Z_{norm}}$ becomes negligible when the number of training samples reaches $\sim$5,000. The mean absolute error and the root error also show a very small dcrease beyond 5,000 training samples, but the decrease is more substantial compared to the $\overline{\Delta Z_{norm}}$. Therefore, more than 5,000 training samples will make a minor contribution to the $\overline{\Delta Z_{norm}}$, while having somewhat higher impact on the mean absolute error and the root error. It should be noted that for the machine learning algorithms used in this study, using large training sets as shown in the table does not add substantial computing requirements.

\begin{table}
\caption{Statistics for various training set sizes. The \texttt{KBestMod} feature set was used with the random forest algorithm}
\label{tab:training-size}
\begin{tabular}{|l|l|l|l|}
\hline
\textbf{Training Set Size} & \textbf{Mean Abs. Err.} & \textbf{Root Err.} & \textbf{$\overline{\Delta Z_{norm}}$} \\ 
\hline
1000              & 0.00836        & 0.02027        & 0.00266  \\ 
5000              & 0.00646        & 0.01446        & 0.00219  \\ 
10000             & 0.00626        & 0.01342        & 0.00221  \\ 
20000             & 0.00617        & 0.01294        & 0.00222  \\
\hline
\end{tabular}
\end{table}

\section{Catalog}
\label{catalog}

The purpose of the photometric redshift methods described in Section~\ref{methods} is to compute the photometric redshifts of the objects in the catalog of galaxy morphologies described in Section~\ref{data}. The photometric redshift of the galaxies in the catalog was computed with the \texttt{KBest\_mod} feature set and the random forest algorithm. 

The catalog contains the 2,912,341 SDSS galaxies classified automatically to spiral and elliptical galaxies \citep{kuminski2016computer}. For each galaxy, the catalog contains the SDSS DR8 object ID of the galaxy, its right ascension, declination, elliptical and spiral marginal probabilities, and the computed photometric redshift. 

\begin{figure}
  \centering
  \includegraphics[scale=0.65]{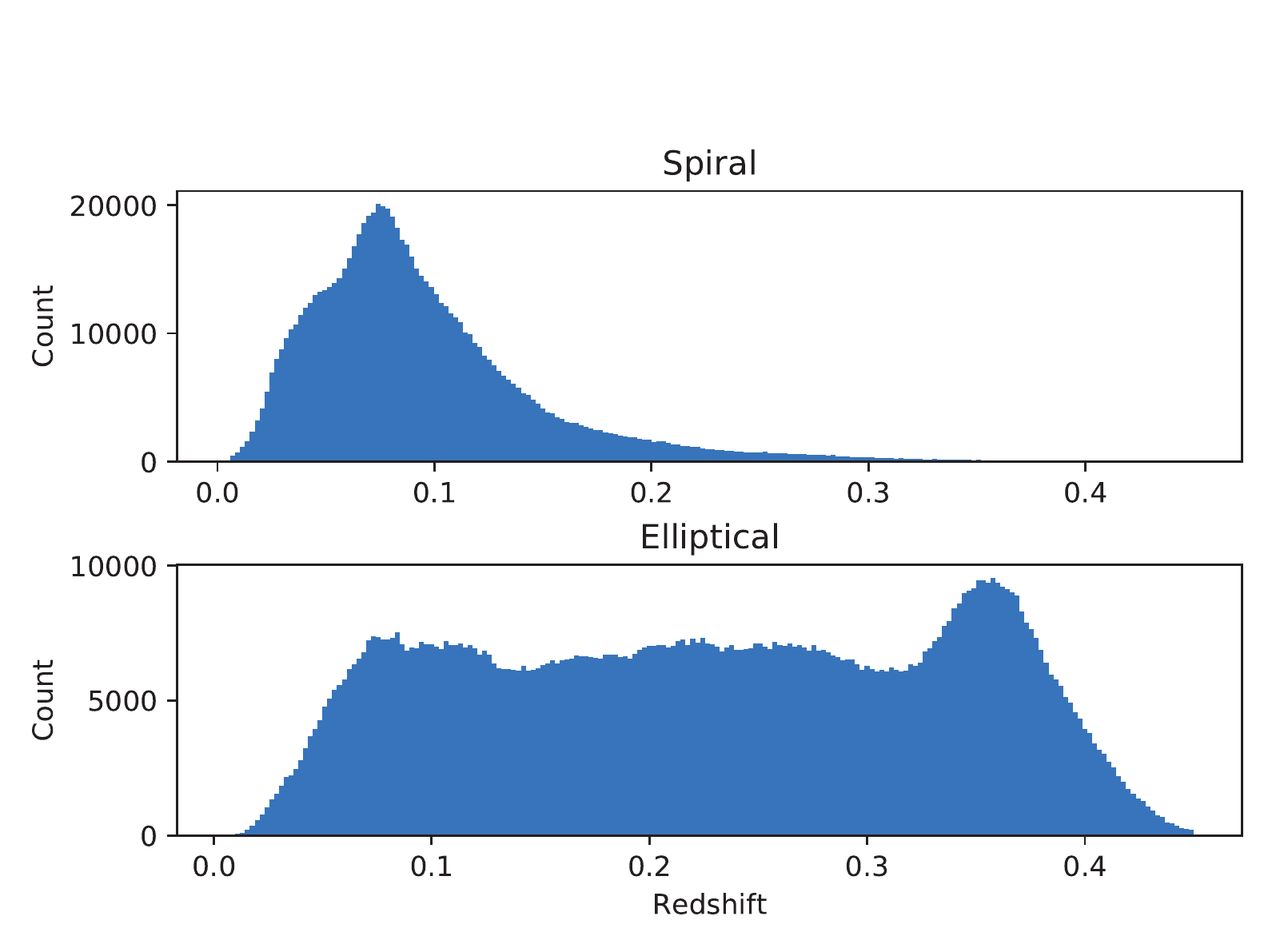}
  \caption{Redshift distribution of the elliptical and spiral galaxies in the catalog.}
  \label{redshift_distributions}
\end{figure}

Figure \ref{redshift_distributions} displays the distribution of the galaxies in the catalog across different photometric redshift ranges. The figures shows that the number of elliptical galaxies remains fairly constant across the redshift ranges, but increases at around redshift of 0.35. The higher number of galaxies in that redshift range is aligned with previous studies, showing a peak in the total number of galaxies at around z=0.35 \citep{zaninetti2015number}. On the other hand, it should be noted that the drop in the number of elliptical galaxies beyond z=0.35 can be related to the limiting magnitude of the catalog. Galaxies with i magnitude dimmer than 18 are excluded from the catalog, and the number of galaxies with redshift greater than 0.35 that satisfy the magnitude threshold is small, and gets smaller as the redshift increases and consequently the galaxies get dimmer. The number of spiral galaxies peaks at around redshift of 0.085, and then decreases gradually.

\begin{figure}
  \centering
  \includegraphics[scale=0.65]{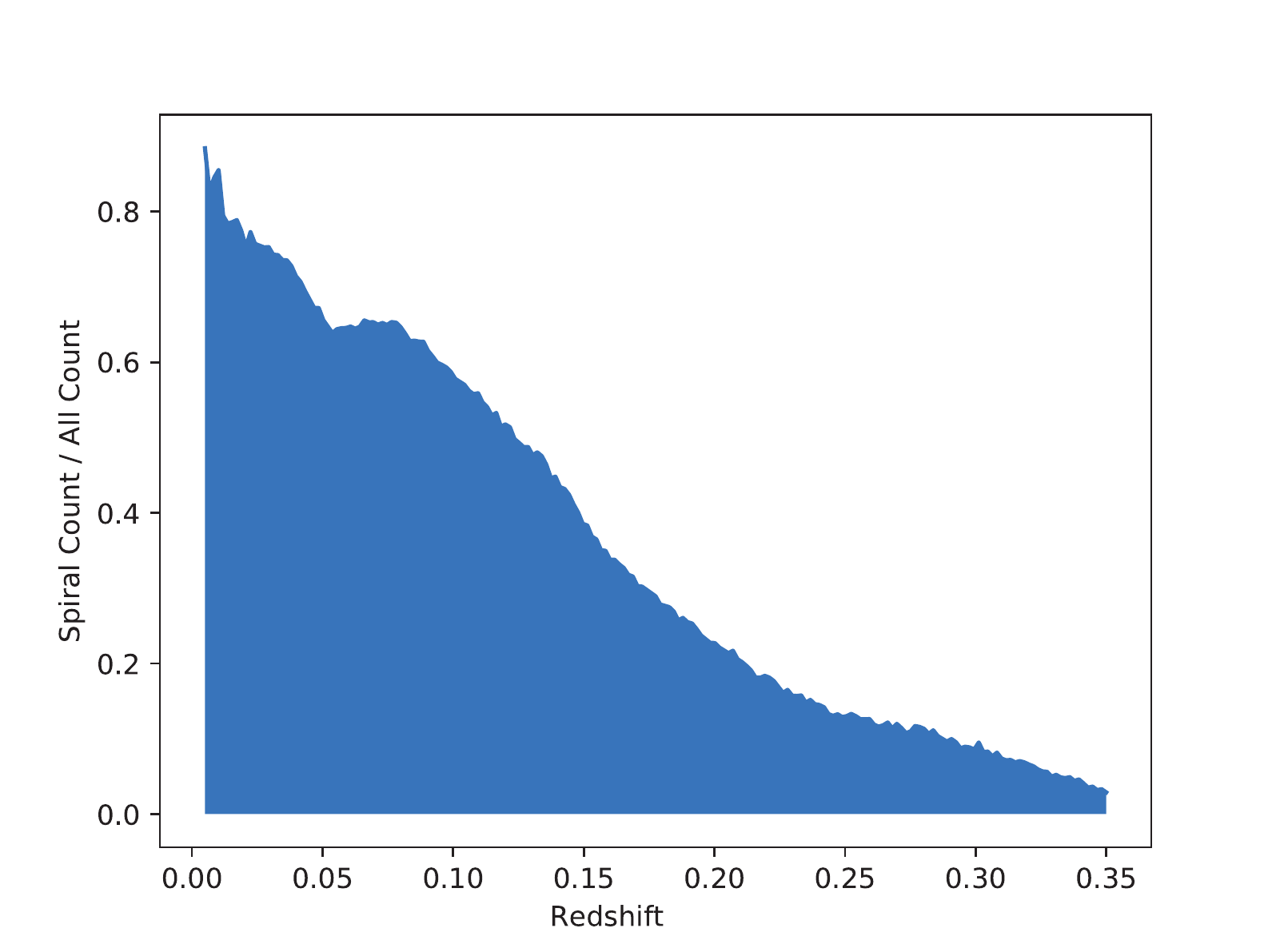}
  \caption{Ratio of spiral galaxies over all galaxies in the catalog in different redshift ranges}
  \label{fig:spiral-ratio}
\end{figure}

The catalog was also analyzed regarding the distribution of broad morphology of the galaxies in each redshift range. Figure~\ref{fig:spiral-ratio} shows the fraction of galaxies identified as spiral among the total number of galaxies within each photometric redshift range. As the figure shows, the proportion of spiral galaxies in the catalog drops as the redshift gets higher. In the redshift range of 0.05-0.1 the fraction of spiral galaxies in the total number of galaxies is $\sim0.62$, while it is $\sim$0.4 at the redshift of 0.15, and drops to less than 0.2 when the redshift is 0.2 or higher. 

It should be noted that the population of galaxies in the catalog might not represent a random sample of the galaxies in the local universe, but is limited to galaxies that are sufficiently bright and sufficiently large to be analyzed morphologically given the limitation of SDSS imaging power. Elliptical galaxies are brighter than spiral galaxies \citep{tempel2011galaxy}, and therefore the magnitude threshold (i magnitude $<$18) can lead to the higher number of elliptical galaxies that meet that threshold to be included in the catalog. Because the redshift and magnitude are strongly correlated, at higher redshifts more elliptical galaxies pass the magnitude threshold compared to spiral galaxies, consequently leading to the higher population of elliptical galaxies compared to spiral galaxies at these redshift ranges.

The increased population of elliptical galaxies at higher redshifts can also be the result of the fact that spiral patterns become more difficult to identify in fainter and smaller galaxies, although the galaxies are all relatively large (Petrosian radius larger than 5.5'') and bright (i magnitude $<$18), and the annotations of the morphologies of these galaxies agree to a very high extent of $\sim$98\% with the debiased ``superclean'' annotations of Galaxy Zoo. Because the initial selection of galaxies is based on the magnitude, more elliptical galaxies with higher photometric redshift can be included in the catalog, and therefore the increase of their population at higher redshifts does not necessarily reflect higher population in the higher redshift ranges of this catalog. On the other hand, the size threshold ensures that only large objects are selected, so that bright and small objects are excluded from the catalog.



\begin{figure}
 \centering
 \includegraphics[scale=1.0]{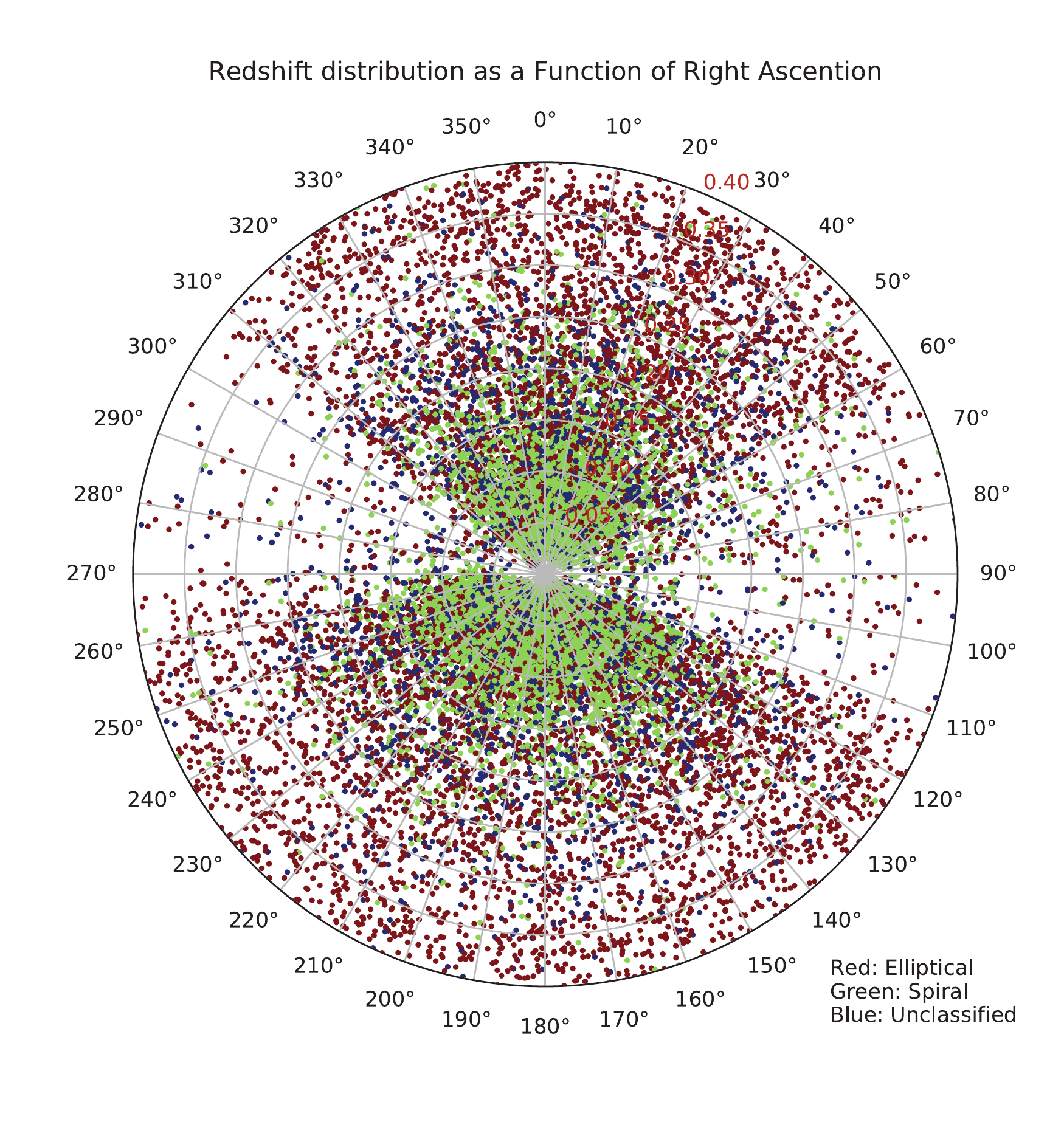}
 \caption{Distribution of spiral and elliptical galaxies by photometric redshift and right ascension. The range of the photoz is 0 to 0.4.}
 \label{zphot_distribution}
\end{figure}

Figure~\ref{zphot_distribution} shows the distribution of the galaxies in the catalog by their broad morphology, right ascension, and photometric redshift. Similarly to Figure~\ref{fig:spiral-ratio}, the lower redshift range has a much higher number of spiral galaxies, which can be also related to the fact that spiral patterns are more difficult to identify as the redshift gets higher.

The redshift range of the galaxies used in the catalog is relatively small in terms of galaxy evolution. Some observations within that redshift range have been noted, such as the higher population of faint blue galaxies at redshift range of 0.3 to $\sim$1 \citep{lilly1996canada}. A more recent observation showed that the population of settled disk galaxies changes in the range of $0.2<z<1.2$ \citep{kassin2012epoch}. The absolute magnitude of galaxies also increases (becomes dimmer) when the redshift increases in the range of $0<z<1$ \citep{zaninetti2015number}. Another study showed a decrease in the population of massive late type galaxies in the $0<z<0.35$, which largely agrees with the distribution of the late-type galaxies in this catalog  \citep{conselice2014evolution}.





\section{Conclusion}

The primary goals of this study are to test machine learning and variable selection algorithms for computing photometric redshift, optimize it for a specific population of galaxies, and mainly apply these algorithms to provide a large catalog of galaxy morphology and photometric redshift. 


The catalog presented in this paper is similar to the early Galaxy Zoo 1 catalog, but because it was classified automatically it provides a much higher number of galaxies. Of the $\sim3\cdot10^6$ galaxies in the catalog, $\sim1.5\cdot10^6$ are galaxies with 98\% agreement rate with the Galaxy Zoo 1 debiased ``superclean'' accuracy. It is limited in the sense that, like Galaxy Zoo, it represents the galaxies in the catalog, and not necessarily a complete and unbiased sample of SDSS galaxies.

The catalog is publicly available at \url{https://figshare.com/articles/Morphology_and_photometric_redshift_catalog/4833593}. The source code used to create the catalog is also publicly available \citep{shamir2008wndchrm,shamir2013wnd}.

\section{Acknowledgments}

This study was supported by NSF grant IIS-1546079. Funding for the SDSS and SDSS-II has been provided by the Alfred P. Sloan Foundation, the Participating Institutions, the National Science Foundation, the US Department of Energy, the National Aeronautics and Space Administration, the Japanese Monbukagakusho, the Max Planck Society, and the Higher Education Funding Council for England. The SDSS Web Site is http://www.sdss.org/. The SDSS is managed by the Astrophysical Research Consortium for the Participating Institutions. The Participating Institutions are the American Museum of Natural History, Astrophysical Institute Potsdam, University of Basel, University of Cambridge, Case Western Reserve University, University of Chicago, Drexel University, Fermilab, the Institute for Advanced Study, the Japan Participation Group, Johns Hopkins University, the Joint Institute for Nuclear Astrophysics, the Kavli Institute for Particle Astrophysics and Cosmology, the Korean Scientist Group, the Chinese Academy of Sciences (LAMOST), Los Alamos National Laboratory, the Max Planck Institute for Astronomy (MPIA), the Max Planck Institute for Astrophysics (MPA), New Mexico State University, Ohio State University, University of Pittsburgh, University of Portsmouth, Princeton University, the United States Naval Observatory and the University of Washington.

\appendix
\section{A: Variable description}

Table~\ref{variable_description} provides a short description of the variables used in the different experiments. A more detailed description is available in the Sloan Digital Sky Survey documentation \citep{aihara2011eighth}. {\bf The variables that were used for the photometric redshift are only the variables included in the KbestMod feature set described in Section~\ref{variable_selection}, and not the entire list of variables described in the table.}

\begin{table}[ht]
\caption{Variables used for photometric redshift estimation. The variables used for the photometric redshift are the subset of variables included in the KbestMod feature set, and not the entire set of variables included in the table.}
\label{variable_description}
\begin{tabular}{|l|l|}
\hline
Variable & Description \\ 
\hline
u     &  The best of the exponential fit magnitude and the DeVaucouleurs fit magnitude in the u band \\ 
g    &   The best of the exponential fit magnitude and the DeVaucouleurs fit magnitude in the g band     \\ 
 r    &  The best of the exponential fit magnitude and the DeVaucouleurs fit magnitude in the r band      \\ 
i    &    The best of the exponential fit magnitude and the DeVaucouleurs fit magnitude in the i band    \\ 
z    &   The best of the exponential fit magnitude and the DeVaucouleurs fit magnitude in the z band     \\ 
deVMag\_u    &   DeVaucouleurs fit magnitude in the u band     \\ 
deVMag\_g    &   DeVaucouleurs fit magnitude in the g band     \\ 
deVMag\_r    &   DeVaucouleurs fit magnitude in the r band     \\ 
deVMag\_i    &    DeVaucouleurs fit magnitude in the i band    \\ 
deVMag\_z    &  DeVaucouleurs fit magnitude in the z band      \\ 
fiberMag\_u    &  Magnitude measured in 3 arcsec diameter fiber radius in the u band \\ 
fiberMag\_g    &   Magnitude measured in 3 arcsec diameter fiber radius in the g band \\ 
fiberMag\_r    &  Magnitude measured in 3 arcsec diameter fiber radius in the r band \\ 
fiberMag\_i    &    Magnitude measured in 3 arcsec diameter fiber radius in the i band\\ 
fiberMag\_z    &   Magnitude measured in 3 arcsec diameter fiber radius in the z band \\ 
expMag\_u    &    Exponential magnitude in the u band \\ 
expMag\_g    &    Exponential magnitude in the g band \\ 
expMag\_r    &     Exponential magnitude in the r band \\ 
expMag\_i    &  Exponential magnitude in the i band \\ 
expMag\_z    &   Exponential magnitude in the z band \\ 
petroMag\_u    &  Petrosian magnitude in the u band \\ 
petroMag\_g    & Petrosian magnitude in the g band \\ 
petroMag\_r    &  Petrosian magnitude in the r band \\ 
petroMag\_i    &   Petrosian magnitude in the i band \\ 
petroMag\_z    &   Petrosian magnitude in the z band \\ 
psfMag\_u    &   PSF magnitude in the u band \\ 
psfMag\_g    &  PSF magnitude in the g band  \\ 
psfMag\_r    &  PSF magnitude in the r band \\ 
psfMag\_i    & PSF magnitude in the i band \\ 
psfMag\_z    & PSF magnitude in the z band  \\ 
psfMagErr\_u    &  PSF magnitude error in the u band \\ 
psfMagErr\_g    & PSF magnitude error in the g band  \\ 
psfMagErr\_r    &  PSF magnitude error in the r band \\ 
psfMagErr\_i    &  PSF magnitude error in the i band \\ 
psfMagErr\_z    & PSF magnitude error in the z band  \\ 
extinction\_u    &   extinction in the u band \citep{schlegel1998maps}     \\ 
extinction\_g    &   extinction in the g band  \\ 
extinction\_ r   &    extinction in the r band  \\ 
extinction\_i    &    extinction in the i band   \\ 
extinction\_z    &   extinction in the z band \\ 
photozcc2 & CC2 algorithm estimation \\
photozd1 & D1 algorithm estimation \\
dered\_u    &  Model magnitude subtracted by the extinction in the u band \\ 
dered\_g    &  Model magnitude subtracted by the extinction in the g band \\ 
dered\_r    & Model magnitude subtracted by the extinction in the r band \\ 
dered\_i    & Model magnitude subtracted by the extinction in the i band \\ 
dered\_z    &  Model magnitude subtracted by the extinction in the z band \\ 
isoA\_u    &  Isophotal major axis in the u band \\ 
isoB\_u    &  Isophotal minor axis in the u band \\ 
isoBGrad\_u    &  Gradient in minor axis in the u band \\ 
isoCocl\_u    &  Isophotal column centroid in the u band \\ 
isoPhi\_u    &   Isophotal position angle in the u band \\ 
isoRowcGrad\_u    &   Gradient in row centroid in the u band \\ 
\hline
\end{tabular}
\end{table}

\reftitle{References}
\externalbibliography{yes}
\bibliographystyle{mdpi}
\bibliography{main}

\end{document}